\documentclass[pra,showpacs,showkeys, sort&compress, floatfix, amsfonts,square]{revtex4}

\usepackage[dvips]{graphicx}
\usepackage{bm}

\newcommand{\newc}{\newcommand}
 
\newc{\be}{\begin{equation}}
\newc{\ee}{\end{equation}}
\newc{\ba}{\begin{eqnarray}}
\newc{\ea}{\end{eqnarray}}
\newc{\bea}{\begin{eqnarray}}
\newc{\eea}{\end{eqnarray}}
\newc{\D}{tial}
\newc{\ie}{{\it i.e.} }
\newc{\eg}{{\it e.g.} }
\newc{\etc}{{\it etc.} }
\newc{\etal}{{\it et al.} } 
\newc{\ra}{\rightarrow}
\newc{\lra}{\leftrightarrow}
\newc{\lsim}{\buildrel{<}\over{\sim}}
\newc{\gsim}{\buildrel{>}\over{\sim}}

\begin{document}

\title{Magnetic domain-walls and the relaxation method}

\author{C. Tannous and J. Gieraltowski}
\affiliation{Laboratoire de Magn\'etisme de Bretagne - CNRS FRE 2697\\
Universit\'e de Bretagne Occidentale -\\
6, Avenue le Gorgeu C.S.93837 - 29238 Brest Cedex 3 - FRANCE}

\begin{abstract}
The relaxation method used to solve boundary value problems 
is applied to study the variation of the magnetization
orientation in several types of domain walls that occur in
ferromagnetic materials. The algorithm is explained and applied
to several cases: the Bloch wall in bulk magnetic systems, the 
radial wall in cylindrical wires  and the N\'eel wall in thin films.  

\end{abstract}

\pacs {75.60.Ch; 75.70.Kw; 02.60.Cb; 02.60.Lj }

\keywords {Domain walls and domain structure; Domain structure; Numerical simulation, 
solution of equations; Ordinary and partial differential equations, boundary value problems}

\maketitle

\section{Introduction}

\begin{figure}[!ht]
\begin{center}
\scalebox{0.5}{\includegraphics*[angle=0]{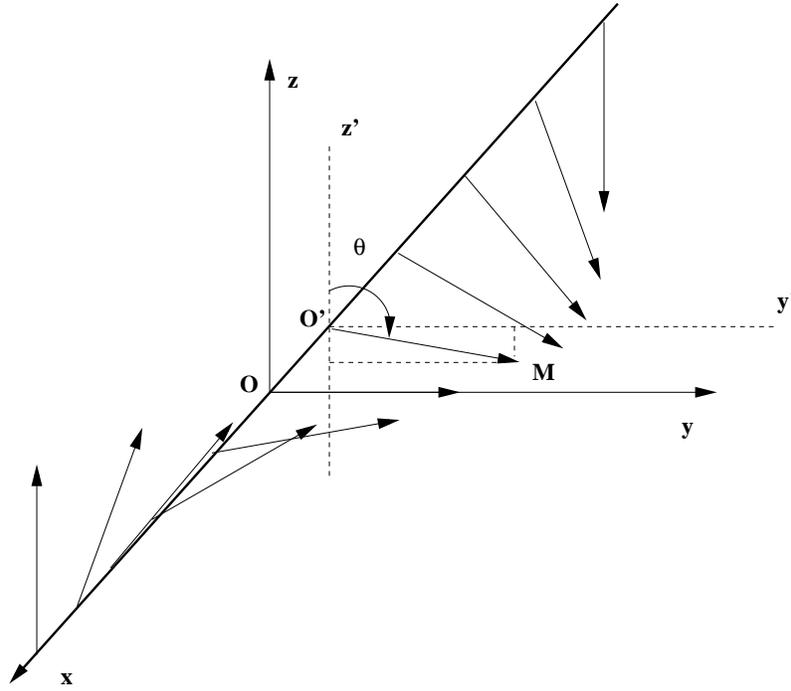}}
\end{center}
  \caption{Behaviour of the magnetization direction for a Bloch wall. For an arbitrary
point along the $x$ axis, the magnetization $\bm{M}$ whose rotation is entirely confined 
within the vertical $zOy$ plane makes the angle $\theta$ with the vertical $z$ axis, the anisotropy
axis.}
\label{bwall}
\end{figure}

A domain is a region in a ferromagnetic material with the magnetization along a given 
direction. A magnetic material contains many domains with  different
magnetizations pointing in different directions in order to minimize the 
total magnetostatic  energy. Regions with different orientations of their 
magnetization can be close to one another albeit with a boundary called a domain wall
(containing typically about 10$^{2}$ -- 10$^{3}$ atoms). \\
Saturation occurs when all these regions align along some common direction 
imposed an external applied field, the total magnetization
reaching its largest value $M_s$. \\
 The width of a domain wall is equal  to $\pi \sqrt{A/K}$ where $A$ is the typical 
nearest neighbor Heisenberg exchange interaction and $K$ the typical anisotropy
constant (see Table~\ref{ferro}). Hence, a magnetic wall results from exchange and anisotropy, being thinner for 
higher anisotropy or smaller exchange (In Fe it is about 30 nanometers whereas in a 
hard material like Nd$_{2}$Fe$_{14}$B it is about 5 nanometers, only). 
Domain wall energy is given by $4\sqrt{AK}$ illustrating once again
the competing role of exchange and anisotropy. \\
For bulk materials, walls of the Bloch type occur whereas in thin films N\'{e}el type walls are
encountered when the film thickness is close to the exchange length (defined by $\ell_{ex}=\sqrt{A/K}$,
which is a few nanometers for ferromagnetic materials like Ni, Fe or Co, see Table~\ref{ferro}).
In the case of soft or amorphous materials characterised by a vanishing 
anisotropy constant $K$, one uses rather the magnetostatic 
exchange length defined by $\ell_{ex}= \sqrt{A/M_s^2}$. In all cases, the wall width $\delta$
is obtained from the exchange length via  $\delta= \pi \ell_{ex}$. \\
A single parameter $Q=2K/M_{s}^{2}$ allows to discriminate between simple 
($Q<1$) and complex wall profiles $(Q>1)$ (see Malozemoff and Slonczewski \cite{malo}). 
For example, in fig.~\ref{bwall} a Bloch wall, belonging to the class ($Q<1$)
is depicted with the magnetization rotating in a vertical plane. \\
Mathematically, a domain wall appears as a result of a non-linear two-point boundary
value problem (TPBVP) since it separates two distinct regions with a well
defined value of the magnetization. The TPBVP originates from a minimization
of the total magnetic energy that contains in general a competition between
the anisotropy and exchange energies. \\
In this work, a general numerical approach based on the relaxation method 
is applied to the study of domain profiles in several geometries: bulk,
wires and thin films. \\
This report is organised as follows: In section 2, the numerical relaxation
method is described; in section 3 we discuss Bloch walls, whereas
radial walls in cylindrical wires are described in section 4.
In section 5 N\'eel walls are described and finally section 6 contains
a discussion and a conclusion.

\section{The relaxation Method}
\label{der}
Traditionally, TPBVP are typically tackled with the shooting method.
The shooting method typically progresses from one boundary point to another
using, for instance, Runga-Kutta integration \cite{recipes} with a set of initial
conditions attempting at reaching the end boundary. \\
For regular Ordinary Differential Equations (ODE), simple shooting is
enough to reach the solution. In more complicated ODE, one has to
rely on double shooting  also called shooting to a fitting point.
The algorithm consists of shooting from both boundaries to a middle
point (fitting point) where continuity of the solution and derivative
are required. In certain cases, one even has to perform 
multiple shooting in order to converge toward the solution \cite{stoer}. \\
In the case of presence of singularities (within the domain or at the
boundaries) the shooting method in all its versions: simple, double or multiple
does not usually converge. We find that it is the case also with domain walls
because of a rapid drop of the solution somewhere in the integration interval (due
to the rapid change of the magnetization orientation in the wall).
In this work, we develop, a new method to tackle the domain wall problem based on 
the relaxation method and find it quite suitable to handle relatively fast changes
in the solution.  \\

The basic idea of the relaxation method is to convert the differential 
equation into a finite difference equation (FDE).  
When the problem involves a system of $N$ coupled first-order ODE's represented 
by FDE's on a mesh of $M $ points, a solution consists of values for $N$ dependent
 functions given at each of the $M$ mesh points, that is $N \times M$ variables in all. 
The relaxation method determines the solution by starting with a guess and 
improving it, iteratively.  
The iteration scheme is very efficient since it is based on the multidimensional 
Newton's method (see {\it Numerical recipes}~\cite{recipes}).
The matrix  equation that must be solved, takes a special,  block diagonal 
 form, that can be inverted far 
more economically both in time and storage than would be possible for a general 
matrix of size $(MN) \times (MN)$. 
The solution is based on error functions for the boundary conditions
and the interior points. \\
Given a set of $N$ first-order ODE's depending on a single spatial variable $x$:
\be
\frac{\bm{dy}_j}{dx}= \bm{g}_j(x,\bm{y}_{1},\dots,\bm{y}_{N}), j=1,2, \dots N
\ee
we approximate them by the algebraic set:
\be
0= \bm{E}_{k} = \bm{y}_k- \bm{y}_{k-1}-(x_{k}-x_{k-1})\bm{g}_k(x_{k},x_{k-1},\bm{y}_{k},\bm{y}_{k-1}),
k=2,3, \dots M
\ee
over a set of  $M-1$ mesh points defining  $[x_{k-1},x_{k}]$ intervals with $k=2,3, \dots M$. \\
The FDE $\bm{E}_{k} $ provide $N$ equations coupling $2N$ variables at the 
mesh points of indices $k-1,k$. The FDE's provide
a total of $M(N-1)$ equations for the $MN$ unknowns.
The remaining equations come from the boundary conditions \cite{recipes}: \\ 
At the first boundary $x_1$ we have: $0= \bm{E}_1 = \bm{B}(x_1,\bm{y}_1)$  \\ 
At the second boundary $x_2$, we have: $0= \bm{E}_{M+1} = \bm{C}(x_M,\bm{y}_M)$  \\
The vectors $ \bm{E}_1$ and $\bm{B}$ have $n_1$ non-zero components corresponding
to the $n_1$ boundary conditions at $x_1$.
The vectors $ \bm{E}_{M+1}$ and $\bm{C}$ have $n_2$ non-zero components corresponding
to the $n_2$ boundary conditions at $x_2$, with $n_1+n_2=N$ the total number of ODE's. \\

The main idea of the relaxation method
is to begin with initial guesses of $\bm{y}_{j}$ and relax them to the
approximately true values by calculating the errors $\bm{E}_{i}$ to correct the value of
$\bm{y}_{j}$ iteratively.  Relaxation might be viewed 
as a rotation of the initial vector (representing the solution)  under the 
constraints defined by $\bm{E}_{i}$.
The evolution of the relaxation process, is obtained from solution-improving 
increments $\Delta \bm{y}_k$  that can be evaluated from a first-order Taylor expansion
of the error functions  $\bm{E}_{k}$. \\
It is that expansion that results in  the matrix equation possessing a special
block diagonal form, allowing inversion economically in terms of time 
and storage (see ref.~\cite{recipes}).

\section{Bloch walls}

The energy of an uniaxial ferromagnetic material comprises 
anisotropy and exchange terms. An infinite volume is
considered to exclude any shape related demagnetization energy.
Exchange energy density is given by \cite{landau}:
\be 
\frac{A_{ik}}{2} (\frac{\partial M_l}{\partial x_i})(\frac{\partial M_l}{\partial x_k})
\ee
where Einstein summation convention is used for repeated indices $i,k,l=1...3$.
The uniaxial anisotropy energy is given by $K_{ij}M_i M_j$ with $i,j=1...3$.
For simplicity, we assume a single uniform exchange constant $A$ (see Table~\ref{ferro})
and a sole dependence on the $x$ coordinate of all components of the magnetization $\bm{M}$. We have
$\bm{M}= (0, M_s \sin\theta(x), M_s \cos \theta(x))$ (see fig.~\ref{bwall}).
$\theta(x)$, the angle the magnetization makes with the $z$ axis considered as
the anisotropy axis. The sought profile is the function $\theta(x)$. 
$M_s$ is the saturation magnetization when all individual
magnetic moments in the material are aligned along the same direction. \\

Integrating  over all the volume, the total energy is given by:

\be
E=\int_{-\infty}^{\infty}\{\frac{A}{2}{(\frac{\partial \bm{M}}{\partial x})}^2
 + \frac{K}{2}M_y^2 \}dx
\ee

This can be rewritten as:

\be
E= \frac{M_s^2}{2}\int_{-\infty}^{\infty} [A{(\frac{d\theta}{dx})}^2+ K \sin^2 \theta ]dx
\ee

The energy minimum is found by nulling the variational derivative of $E$ with
respect to $\theta$. We find:
 
\be
\frac{d^2\theta}{dx^2}-\xi \sin \theta \cos \theta=0
\ee

with $\xi=\frac{K}{A}$. The Bloch wall profile is given by the solution to the above 
second-order ODE written as a system of two first-order equations:

\ba
\frac{dy_1}{dx} & = & y_2  \\
\frac{dy_2}{dx} & = & \xi \sin \theta \cos y_1
\ea

where $y_1=\theta(x)$ satisfies the boundary conditions:

\be
 \lim_{x\rightarrow -\infty} y_1(x) = \pi;  \hspace{1cm}  \lim_{x\rightarrow \infty} y_1(x) = 0
\ee

It is understood that the sharp transition between the $\theta=0$ phase and the
$\theta=\pi$ is behind the failure of all shooting methods.

\begin{figure}[!ht]
\begin{center}
\scalebox{0.6}{\includegraphics*[angle=0]{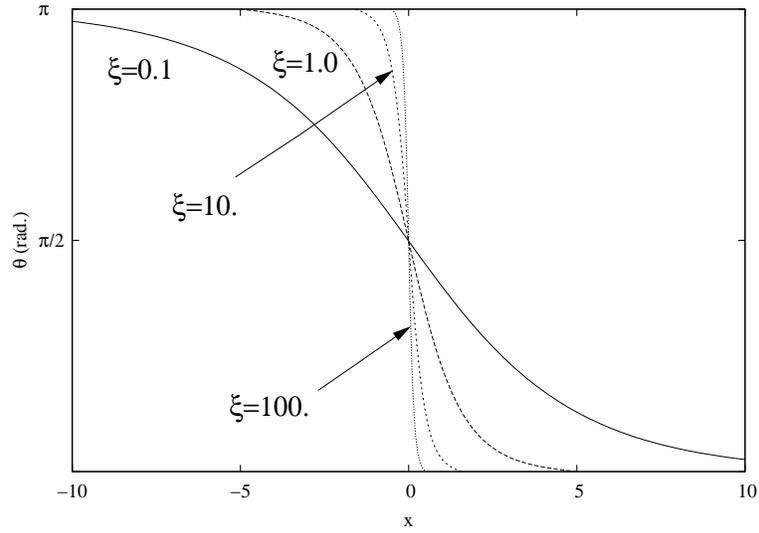}}
\end{center}
  \caption{Variation of the magnetization angle with distance for a Bloch wall for
various values of the exchange anisotropy ratio $\xi$.
The analytical result $\theta(x)=2 \tan^{-1}(e^{-x})$, for $\xi=1$, is
indistinguishable from the relaxation method result.}
\label{bloch}
\end{figure}

Using the relaxation method, we easily obtain the wall profiles for any value of $\xi$
as displayed in fig.~\ref{bloch}. 

Actually, the Bloch wall problem is single scale and we can without performing the calculation
for every $\xi$ value, do it for one value and then change the scale accordingly.
This is done as follows:  The width of the domain wall is 
given by $\delta = 1/\sqrt{\xi}$ as explained previously.
We perform a scaling transformation to the $x$ coordinate as: $\tilde{x}=x/\delta$ turning
the ODE into:

\be
\frac{d^2\theta}{d\tilde{x}^2}- \sin \theta \cos \theta=0
\ee

The exact  analytical solution of the above equation given by: 
$\theta(\tilde{x})=2 \tan^{-1}(e^{-\tilde{x}})$ is indistinguishable from the relaxation
method results displayed in fig.~\ref{bloch}.

Numerically, this means one can do the calculation for $\xi=1$ and later on rescale the $x$ variable 
in order to get the solution for any value (arbitrarily large or small) of $\xi$. 
Despite the power of the relaxation method, we noticed that when  $\xi \le 10^{-5}$ or
when  $\xi \ge 10^{4}$, convergence becomes difficult due to rounding and conditioning 
errors.

That rescaling works for many types of walls except
N\'eel wall where we have an additional scale controlling the profile (see section V).
         
\section{Radial walls}

\begin{figure}[!ht]
\begin{center}
\scalebox{0.5}{\includegraphics*[angle=0]{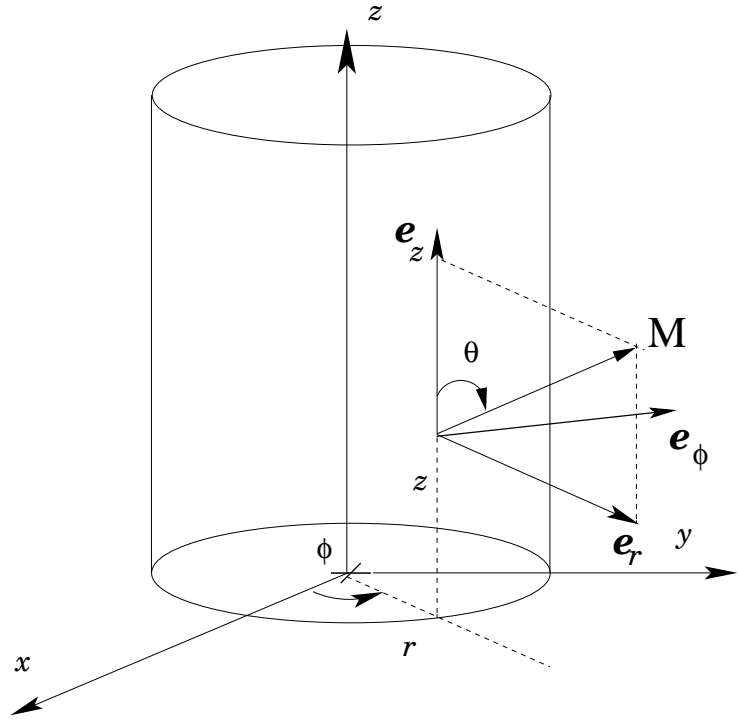}}
\end{center}
  \caption{Cylindrical coordinates displaying the spatial variation of the magnetization 
angle with radial distance from the wire axis. The base plane perpendicular to the wire
axis $z$ is an easy plane.}
\label{cylinder}
\end{figure}

The energy density of an infinite cylindrical (see fig.\ref{cylinder})
uniaxial ferromagnetic material comprising uniaxial anisotropy and exchange terms is 
given by:

\be
\frac{A}{2}[{(\frac{\partial \bm{M}}{\partial r})}^2
+ {(\frac{\partial \bm{M}}{r \partial \phi})}^2
+ {(\frac{\partial \bm{M}}{\partial z})}^2] 
\ee

For simplicity, we assume a single uniform exchange constant $A$ and a sole dependence
on the radial coordinate  $r$ of all the magnetization components of $\bm{M}$.
Integrating  over all a cylindrical volume of radius $R$, the total energy is given by:

\be
E=\frac{1}{\pi R^2}\int_{a/2}^{R}\{\frac{A}{2}[{(\frac{d\theta}{dr})}^2 +\frac{\sin^2 \theta}{r^2}]
+ K \cos^2 \theta \} 2\pi rdr
\ee
where $\theta$ is the angle the magnetization makes with the $z$ axis (see fig.~\ref{cylinder}).
$a$ plays the role of a lattice parameter, the minimal core radius, regularising the integral
(see for instance \cite{frei}). As in the Bloch wall case, we consider that $\theta$ is a function
of one spatial coordinate only ($r$ in this case). Since the anisotropy energy is given by: 
 $K \cos^2 \theta$ with $K$ positive, the base plane (perpendicular to the $z$ axis) is easy, meaning
the minimum of anisotropy energy is obtained when $\theta=\pi/2$  (see fig.\ref{cylinder}).

The total energy minimum is found by nulling the variational derivative of $E$ with
respect to $\theta(r)$. We find:
 
\be
\frac{d^2\theta}{dr^2} +\frac{1}{r}\frac{d\theta}{dr}+ 
\frac{\sin 2\theta}{2}[ \xi -  \frac{1}{r^2}]=0
\ee

with $\xi=\frac{2K}{A}$. The radial wall profile is given by the solution to the above 
second-order ODE (equivalent to system of two first-order ODE's like the Bloch case) 
with the boundary conditions:

\be
 \lim_{r\rightarrow a} \theta (r) = 0;  \hspace{1cm}  \lim_{r\rightarrow R} \theta (r) = \pi/2
\ee

The limits: $ a \rightarrow 0; R \rightarrow \infty$ are taken afterwards.

\begin{figure}[!ht]
\begin{center}
\scalebox{0.8}{\includegraphics*[angle=0]{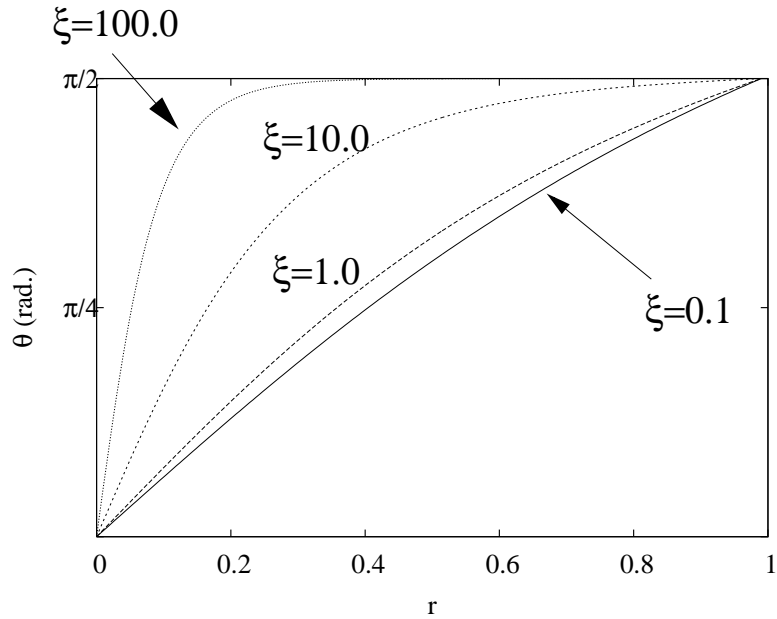}}
\end{center}
  \caption{Variation of the magnetization angle $\theta$ with radial distance $r$ from the wire axis
for various values of the exchange anisotropy ratio $\xi$. As we increase $\xi$ the angle
increases faster from 0 (Easy axis along $z$) to $\pi/2$ (Easy plane $\bot z$)}
\label{radial}
\end{figure}

Using several values of $\xi$ we obtain the radial wall profile in fig.\ref{radial}.
Again, like in the Bloch case, there is a single length involved and it suffices in fact to
solve the TPBVP for a single case $\xi=1$ and rescale all variables accordingly. 
This is not the case of N\'eel walls as decribed in the next section.

\section{N\'eel walls}

N\'eel realized that in a regime where the  
thickness of a ferromagnetic film becomes comparable to the Bloch wall width,
a transition mode within the plane can lower the total energy decisively.
Unlike the Bloch wall problem where only two energy components (exchange and anisotropy) 
exist balanced by a single length scale, the N\'eel wall problem incorporates
two characteristic length scales. The new length arises from the competition 
with an additional energy component, the internal  field energy.  
This has important physical, mathematical and numerical consequences.
On the physical side, a very rich behaviour of N\'eel walls in thin films was shown 
recently in ref.~\cite{garcia}. \\
Domain structures in thin inhomogeneous ferromagnetic films with smooth and small 
inhomogeneities in the exchange and anisotropy parameters yield very complex domain 
structures~\cite{garcia}. Domain walls are fixed near certain inhomogeneities
but do not repeat their spatial distribution. In addition there are metastable chaotic 
domain patterns in periodically inhomogeneous films. \\
The mathematical description of N\'eel walls entails the introduction of an internal
magnetic field $\bm{H}$ created by $\rho$ the 
induced pole density induced by the rotation of $\bm{M}$. Mathematically we
have div$\bm{H}$=-div$\bm{M}$=$\rho$.
The magnetization is expressed as $\bm{M}=(M_s \sin\theta(x),M_s \cos \theta(x),0)$ in the $xyz$ 
coordinates defined in fig.~\ref{nwall}. Since the divergence of $\bm{M}$ is not zero, we
have an induced pole density $\rho$.
In contrast, $\rho=0$ in the Bloch wall case since we
recall in this case (see section III), $\bm{M}= (0, M_s \sin\theta(x), M_s \cos \theta(x)$).
Assuming as done previously, that the components of $\bm{H}$ depend solely on the 
spatial variable $x$ (see fig.~\ref{nwall}), we obtain:
\be 
\rho=-{\rm div}\bm{M}=-\frac{\partial M_x}{\partial x}=-M_s \frac{d(\sin\theta)}{dx}
\label{rho1}
\ee

The ODE that controls the wall profile $\theta(x)$ is derived exactly as before (taking
account of the exchange and anisotropy terms) with the addition of the Zeeman term
accounting for the presence of the internal field $H(x)$:
\be
2A \frac{d^2\theta}{dx^2}-K \sin 2 \theta + M_s H(x) \cos \theta=0
\label{neelODE}
\ee

The uniaxial anisotropy term is $K \sin 2 \theta$ with $\theta$, the angle 
the magnetization makes with the $y$ axis (the anisotropy axis). 

Note that the demagnetization energy (due to the finite thickness of the film
along the $z$ direction) is zero, since it is given by $2\pi N_{ij}M_i M_j$
with $N_{xx}=N_{yy}=0, N_{zz}=1$, and $M_z=0$.

The difference between this equation and the previous ones (Bloch and Radial cases) 
is that the internal field term $\bm{H}$ depends on the profile $\theta(x)$. Writing
$H(x)$ instead of $H(\theta(x))$ makes the wall-profile equations
non-autonomous because of the explicit $x$ dependence in $H(x)$. Additionally
these equations are integro-differential because of the dependence of $H(x)$ on 
$\theta(x)$ (see for instance \cite{cervera}). \\
In this work we consider the thin film approximation and retrieve a system of three ODE's
by introducing a third function  $y_3=H(x)/H_K$ with $H_K=2 K/M_s$ the anisotropy field. 
The ODE system to solve is written with respect to normalised coordinates $\tilde{x}=x/\delta$
where $\delta$ is the wall thickness ($\delta= 1/\sqrt{\xi}$ where, as before,
$\xi=\frac{K}{A}$):

\ba
\frac{dy_1}{d\tilde{x}} & = & y_2  \\
\frac{dy_2}{d\tilde{x}} & = &  \sin y_1 \cos y_1 - y_3 \cos y_1 \\
\frac{dy_3}{d\tilde{x}} & = & -\pi C y_2 \cos y_1 
\label{neelsys}
\ea

The magnitude of the coupling constant $C=\frac{M_s}{\pi H_K}$ has a strong
effect on the solution of the system. In the limit $C=0$ we recover the
simple case with no internal field $H(x)=0$ like the Bloch wall case. 
As the magnitude of $C$ increases, we get
a greater variation in the spatial dependence of $y_3(\tilde{x})$  and the system
might become unstable and display numerical oscillations in spite of a drastic
reduction of the integration step.

We convert the boundary conditions from the $] -\infty, +\infty[$ interval
to the $[0, +\infty[$ interval:
\be
 \lim_{\tilde{x}\rightarrow 0} y_1 (\tilde{x}) = \pi/2;  \hspace{1cm}
  \lim_{\tilde{x}\rightarrow \infty} y_1 (\tilde{x}) = 0  \hspace{1cm} 
 \lim_{\tilde{x}\rightarrow \infty} y_3 (\tilde{x}) = 0
\ee

We describe below a special algorithm, that we developed,
based on the relaxation method coupled to an iterative procedure. 
The pseudo-code follows:
\begin{enumerate}
\item Initially, we introduce a guess profile
(say $\theta_0(x)$), extract from it the pole density $\rho_0(x)$ 
using eq.~\ref{rho1} and determine from it the field derivative using the
divergence equation: $\frac{dH(x)}{dx}=\rho(x)$.
\item The ODE system is solved and that allows us to extract a new profile 
(say $\theta_1(x)$) that yields a new pole density $\rho_1(x)$ (using eq.~\ref{rho1}).
\item We repeat this procedure to the 
$n$-th step with a profile $\theta_n(x)$ yielding a pole density $\rho_{n+1}(x)$ 
that provides a profile $\theta_{n+1}(x)$. The procedure stops when the difference between the
two profiles  $\theta_n(x)$ and $\theta_{n+1}(x)$ in the mean-square sense becomes
smaller than an error criterion.
\end{enumerate}

The latter profile will have then relaxed self-consistently to the
sought optimal profile that minimises the total energy (see ref.~\cite{cervera} 
and references within).

\begin{figure}[!ht]
\begin{center}
\scalebox{1.0}{\includegraphics*[angle=0]{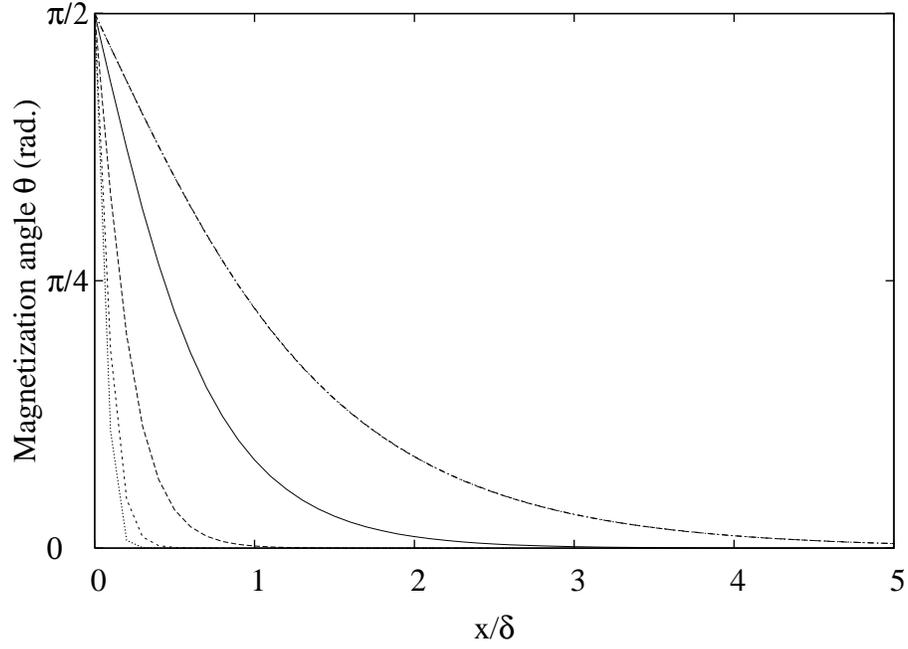}}
\end{center}
  \caption{Variation of the magnetization angle with distance for a N\'eel wall for
various values of the coupling constant $C$. 
Uppermost curve is for $C=0$, whereas lower curves correspond respectively to $C$=1, 10, 50 and finally 100.
The exact result corresponding to zero thickness along the $z$ direction is indistinguishable from the $C=0$ curve.}
\label{neel}
\end{figure}

\begin{figure}[!ht]
\begin{center}
\scalebox{1.0}{\includegraphics*[angle=0]{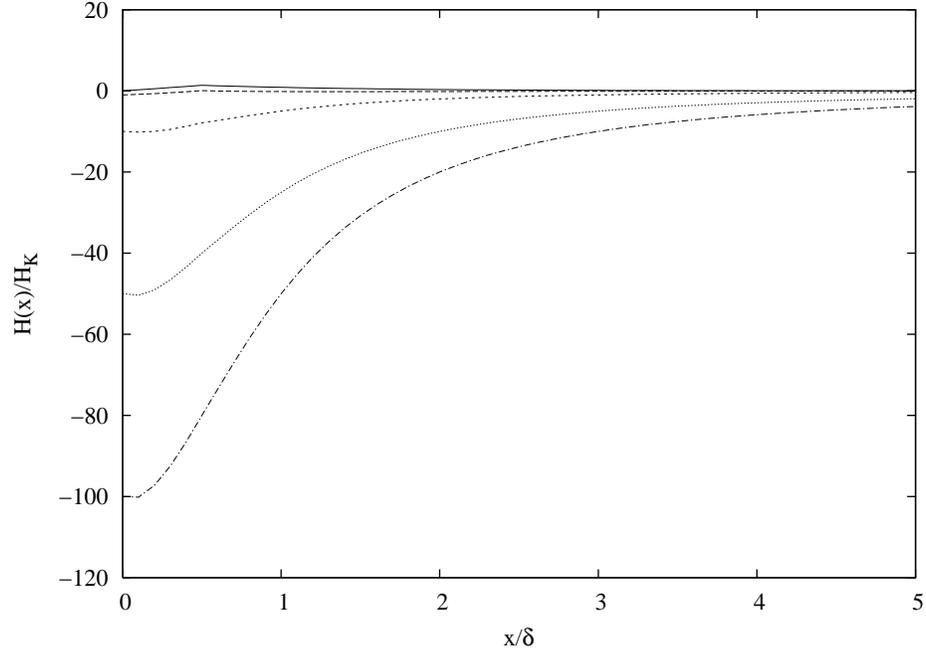}}
\end{center}
  \caption{Variation of the normalised internal field  $H(x)/H_k$ with normalised
distance for a N\'eel wall for various values of the coupling constant $C$. 
Uppermost curve is for $C=0$, whereas lower curves correspond respectively to $C$=1, 10, 50 and finally 100.}
\label{field}
\end{figure}

The results we obtain with various values of $C$ 
for $\theta(\tilde{x})$ and the internal field $H(\tilde{x})/H_K$  are displayed in
fig.~\ref{field} and fig.~\ref{neel}. The analytical result obtained for $C=0$ (Bloch case),
given by: $\theta(\tilde{x})=2 \tan^{-1}(e^{-\tilde{x}})$
is displayed in fig.~\ref{neel} and is indistinguishable from the numerical
results we obtain with the relaxation method on the system~\ref{neelsys}. \\
The results obtained for the internal field displayed in fig.~\ref{field} show, 
as expected (see for instance ref.~\cite{cervera}), that when $C$ increases, the
field (absolute) amplitude becomes larger close to the origin. In addition, as $C$
increases the field extends to larger distances farther from the origin. That, 
in fact, points to the origin of the integro-differential nature of the problem.
Inspection of eq.~\ref{neelODE} shows that in addition to the usual length scale (wall width)
$\delta= 1/\sqrt{\xi}$, we have another length given by: $\delta_N=\sqrt{A/(KC)}=\delta/\sqrt{C}$
arising from the internal field whose strength is given by the coupling constant $C$.
As $C$ increases, non-local effects increase, the length ratio  $\delta_N/\delta=1/\sqrt{C}$
decreases (making the competition between the two lengths harder to deal with because of
the disparity of the two lengths) and it becomes more and more
difficult for the relaxation method to find an optimum result satisfying
the TPBVP.

\section{Discussion and Conclusion}

The magnetic domain profile is a challenging mathematical and numerical problem.
In this work, we treated with the relaxation method, in the simple domain structure case ($Q<1$),
wall configurations in several interesting physical cases: Bloch walls in ferromagnetic
bulk systems, radial walls in cylindrical ferromagnetic wires and the Neel walls in thin
ferromagnetic films assuming in all cases uniaxial anisotropy. \\
In the N\'eel case, we showed than in the  thin film approximation
(in present technology, thin means $\sim 10-100 \AA$) one is able to solve the wall problem with the 
relaxation method with a  proper selection of the variables. Nevertheless, a major 
difficulty appears at higher value of the thickness 
$t$ along the $z$ direction (see fig.~\ref{nwall}). \\
When the thickness of the film increases the system becomes a full integro-differential 
system whereas in the thin film approximation, we get a set of coupled non-linear ODE's that
we have to treat with a special self-consistent algorithm.
The non-locality of the internal field is responsible for the appearance of
logarithmic tails in the spatial variation of the magnetization angle. 
That means the TPBVP must be solved over an ever increasing interval size. 
The algorithm we have developed still applies but one has to use the finite thickness formulas for the
field $H(x)$ (eq.~\ref{fieldeq}) and its derivative $\frac{dH(x)}{dx}$
(eq.~\ref{deriv}) as shown in the Appendix.
The extension of this work to other types of walls (originating from other types
of anisotropy for instance, or the complex wall shape case $Q>1$) or wall dynamics is 
challenging since the wall profile rapid
change imposes a constraint on the time integration step. \\
Previously, Smith treated  domain wall dynamics in small 
patterned magnetic soft thin ($\sim 100 \mu$m) films and turned the dynamic Landau-Lifshitz equations
into a set of coupled non-linear ODE's. It turns out that the system of equations, he found
is stiff (see for instance ref.~\cite{ascher}), imposing a very small integration timestep slowing down considerably the integration process on top of the difficulties of the TPBVP. \\

The extension of this work to domain structures in inhomogeneous media (see ref.~\cite{garcia})
is also quite interesting, 
particularly to the case of thin magnetic films that are of high technological interest such as
recording, memories (Magnetic RAM's and Tunnel Junctions) and Quantum computing and communication. \\

\section{Acknowledgements}
The  authors wish to acknowledge friendly discussions with M. Cormier (Orsay)
regarding dynamic effects in ferromagnetic materials and N. Bertram (San Diego) for sending some of his
papers prior to publication.\\

\begin{center}
{\textbf APPENDIX}
\end{center}

\begin{figure}[!ht]
\begin{center}
\scalebox{0.5}{\includegraphics*[angle=0]{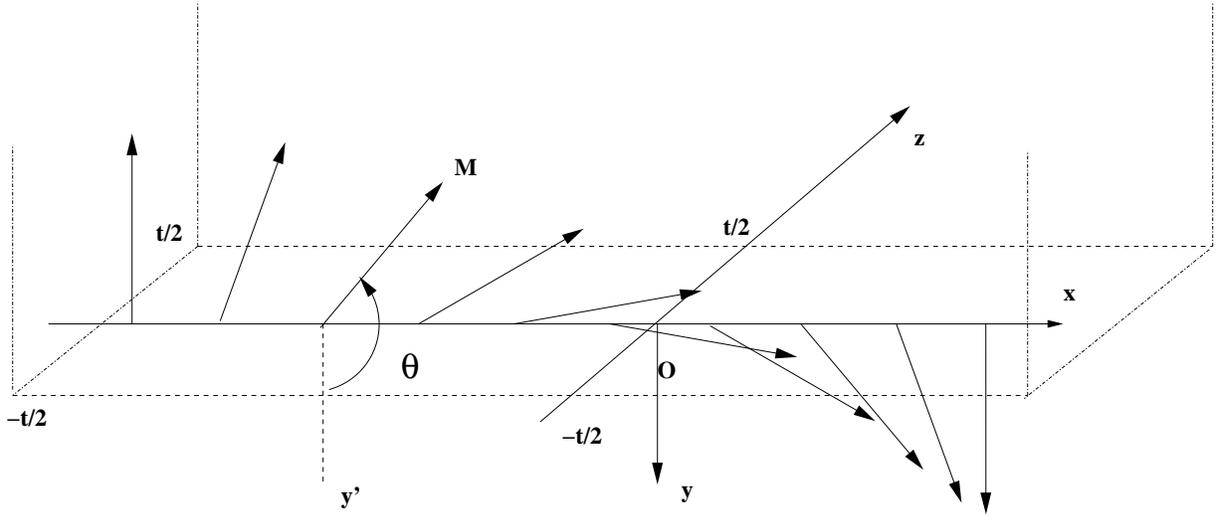}}
\end{center}
  \caption{Variation of the magnetization angle with distance for a N\'eel wall in a thin film. The
angular variation is in the $xOy$ plane. The film is of infinite dimensions along the $x$ and $y$ directions and a finite thickness $t$ along the $z$ direction. The anisotropy axis is along $y$.}
\label{nwall}
\end{figure}

We derive, in this Appendix, the formula for the internal field from the induced pole density.
The magnetization is expressed as $\bm{M}=(M_s \sin\theta(x),M_s \cos \theta(x),0)$ in the $xyz$ 
coordinates defined in fig.~\ref{nwall}. Since the divergence of $\bm{M}$ is not zero, we
have an induced pole density $\rho(x)$. The internal field is obtained from the pole density
by integration accounting for the finite thickness of the film. \\
Using div$\bm{H}=-\rho$ we infer from general theorems of electromagnetism that:

\be
\bm{H(r)}=-\frac{1}{4 \pi}\int d\bm{r'}\rho(\bm{r'})\frac {(\bm{r-r'})}{{|\bm{r-r'}|}^3}
\ee

with $\bm{r}=(x,y,z), \bm{r'}=(x',y',z')$.\\
By symmetry we have $H_y=H_z=0$ and the $x$ component $H(x)$ in the plane $z=0$ is written as:

\be
H(x)=-\frac{1}{4 \pi} \int_{-\infty}^{+\infty}dx'\int_{-\infty}^{+\infty}dy' \int_{-t/2}^{t/2}dz' \rho(x')
\frac {(x-x')}{{[{(x-x')}^2+{(y-y')}^2+{(z')}^2]}^{3/2}}
\ee

A first integration over $y'$ gives:
\be
H(x)=-\frac{1}{2 \pi} \int_{-\infty}^{+\infty}dx' \int_{-t/2}^{t/2}dz' \rho(x')
\frac {(x-x')}{[{(x-x')}^2+{(z')}^2]}
\ee

A second integration over $z'$ yields the result:

\be 
H(x)=\frac{1}{\pi}\int_{-\infty}^{\infty}\rho(x') \tan^{-1}[\frac{t}{2(x-x')}]dx'
\label{fieldeq}
\ee

The relation div$\bm{H}=\rho$ gives the integral expression of $\frac{dH(x)}{dx}$ needed in the
integration of system of ODE's eq.~\ref{neelsys}:
\be 
\frac{dH(x)}{dx}=\frac{2t}{\pi}\int_{-\infty}^{\infty}\rho(x') \frac{1}{4{(x-x')}^2+t^2}dx'
\label{deriv}
\ee

In the finite thickness case, one needs the solve an integro-differential system of equations defined
by the system of ODE's~\ref{neelsys} and the integral eq.~\ref{deriv}. 
In the case of thin films $t \rightarrow 0$, we recover from eq.~\ref{deriv} the previous definition
$\frac{dH(x)}{dx}=\rho(x)$ by using the $\delta$ function definition:

\be 
\delta(x-x')= \frac{1}{\pi} \lim_{t \rightarrow 0} \frac{2t}{4{(x-x')}^2+t^2}
\label{delta}
\ee

\begin{center}
{\textbf  TABLES AND FIGURES}
\end{center}

\begin{table}[htbp]
\begin{center}
\begin{tabular}{ c c c c c c }
\hline
Material & $T_c$  &  $\mu_0 M_s$  &  $A$ &  $K$  & $\ell_{ex}$ \\
\hline
Unit  &  [K]  &  [T]  &  10$^{-11}$[J/m]  &  10$^{5}$ [J/m$^{3}$]  &   [nm] \\
Fe  &  1044  & 2.16  & 1.5  & 0.48    & 2.8  \\
Co  & 1398  & 1.82  & 1.5  &  5  & 3.4  \\
Ni  &  627  &  0.62  &   1.5  &   -0.057  &   9.9  \\
Permalloy   & 720  &  1.0  &   1.3   &  0  &  5.7  \\
CrO$_2$   & 393 & 0.5  &  0.1  &  0.22  &   3.2  \\
SmCo$_5$    & 993  &   1.05  &   2.4  &   170 & 7.4 \\
\hline
\end{tabular}
\caption{Properties of Ferromagnetic Materials: $T_c$ is Curie temperature,
$\mu_0$ is vacuum permeability, $M_s$ is saturation magnetization, $A$ is 
exchange constant, $K$ is magneto-crystalline anisotropy constant and  $\ell_{ex}$ is exchange length.
Note that in the case of Permalloy (Ni$_x$Fe$_{100-x}$ alloys with $x \sim 80$), one uses the magnetostatic 
exchange length defined as $\ell_{ex}= \sqrt{A/M_s^2}$ since $K \sim 0$.}
\label{ferro}
\end{center}
\end{table}

\end{document}